\documentclass[a4paper,11pt]{article}
\pdfoutput=1
\usepackage[intlimits,centertags]{amsmath}
\usepackage{amssymb,amsfonts}
\usepackage{bbm}
\usepackage{mathrsfs}
\usepackage{cite}
\usepackage[scale={0.78,0.8}]{geometry}
\usepackage{multirow}
\usepackage{relsize}
\usepackage[subrefformat=parens,labelformat=parens]{subfig}
\usepackage{xcolor}
\usepackage{caption}
\usepackage{cancel}
\usepackage{graphics,graphicx}
\usepackage{float}
\usepackage{authblk}
\usepackage{lipsum}
\numberwithin{equation}{section}

\def\beq{\begin{align}}
\def\eeq{\end{align}}
\newcommand{\bi}{\begin{itemize}}
\newcommand{\ei}{\end{itemize}}
\newcommand{\ben}{\begin{enumerate}}
\newcommand{\een}{\end{enumerate}}
\newcommand{\be}{\begin{equation}}
\newcommand{\ee}{\end{equation}}
\providecommand{\bea}{\begin{eqnarray}}
\providecommand{\eea}{\end{eqnarray}}

\usepackage[hidelinks]{hyperref}
\usepackage{nicefrac}
\usepackage{booktabs}

\title{\textbf{Stringy multifield quintessence and the Swampland}}

\author[1,2,3]{Max Brinkmann\thanks{maxhorst.brinkmann@unipd.it}}
\author[1,4]{Michele Cicoli\thanks{michele.cicoli@unibo.it}}
\author[5,6]{Giuseppe Dibitetto\thanks{giuseppe.dibitetto@roma2.infn.it}}
\author[1,4]{Francisco G. Pedro\thanks{francisco.soares@unibo.it}}

\affil[1]{\small{Dipartimento di Fisica e Astronomia, Universit\`a di Bologna, via Irnerio 46, 40126 Bologna, Italy}}
\affil[2]{Dipartimento di Fisica e Astronomia, Universit\`a di Padova, via Marzolo 8, 35131 Padova, Italy}
\affil[3]{INFN, Sezione di Padova, via Marzolo 8, 35131 Padova, Italy}
\affil[4]{INFN, Sezione di Bologna, viale Berti Pichat 6/2, 40127 Bologna, Italy}
\affil[5]{\small{Dipartimento di Fisica, Universit\`a di Roma Tor Vergata, via della ricerca scientifica 1, 00133 Rome, Italy}}
\affil[6]{INFN, Sezione di Roma2, via della ricerca scientifica 1, 00133 Rome, Italy}

\date{}

\begin{document}

\maketitle

\begin{abstract}
We consider quintessence models within 4D effective descriptions of gravity coupled to two scalar fields. These theories are known to give rise to viable models of late-time cosmic acceleration without any need for flat potentials, and so they are potentially in agreement with the dS Swampland conjecture. In this paper we investigate the possibility of consistently embedding such constructions in string theory. We identify situations where the quintessence fields are either closed string universal moduli or non-universal moduli such as blow-up modes. We generically show that no trajectories compatible with today's cosmological parameters exist, if one starts from matter-dominated initial conditions. It is worth remarking that universal trajectories compatible with observations do appear, provided that the starting point at early times is a phase of kinetic domination. However, justifying this choice of initial conditions on solid grounds is far from easy. We conclude by studying Q-ball formation in this class of models and discuss constraints coming from Q-ball safety in all cases analyzed here.
\end{abstract}

\tableofcontents

\section{Introduction}
\label{Intro}

The current phase of accelerated expansion observed in our universe poses a surprising challenge for theoretical physicists. On the one hand, the success of simple proposals like the $\Lambda$CDM model \cite{Riess:1998cb,Jaffe:2000tx,Tegmark:2003ud} suggests that a small positive and constant value for the dark energy density (cosmological constant) driving the aforementioned expansion is the best fit \cite{Scolnic:2017caz,Aghanim:2018eyx}, certainly in accord with Occam's Razor. On the other hand, due to the global structure of spacetime, even a free quantum field theory in such a de Sitter (dS) background is rather different from the flat space theories we are used to \cite{Witten:2021jzq}. The main puzzle here being the difficulty of defining an S-matrix that would allow us to consistently formulate a unitary theory of scattering on such a background.

In a theory of quantum gravity like string theory, a cosmological constant must be explained as a positive valued minimum for a scalar field potential. However, while several string constructions which may lead to such a vacuum have been proposed, their reliability still remains highly debated (see \cite{Cicoli:2018kdo} for a systematic analysis of pros and cons of some putative dS string vacua). The difficulty to build explicit metastable dS solutions has even given rise to the common lore that positive minima of scalar potentials might not even exist in quantum gravity \cite{Danielsson:2018ztv}. This statement is commonly referred to as the dS swampland conjecture. More precisely, not only does the dS conjecture forbid the existence of dS extrema but it even restricts the slope of the scalar potentials arising from string compactifications as to prevent shallow profiles in regions where they go positive \cite{Obied:2018sgi}. Further refinements proposed in \cite{Garg:2018reu,Ooguri:2018wrx,Andriot:2018mav} involve second derivatives of the scalar potential. 

An alternative to the cosmological constant scenario to account for late-time cosmic acceleration is to allow for a slowly time-varying dark energy, i.e. a quintessence model (see \cite{Tsujikawa:2013fta} for a review). Similarly to inflation, the motion of a scalar field through its potential drives a phase of accelerated expansion of the universe, its departure from pure dS being due to a non-zero scalar kinetic energy. For this mechanism to work for a single (real) field, in general the corresponding potential must have a very flat profile along the physical trajectory. However \cite{Cicoli:2021fsd} showed that slow roll quintessence cannot be achieved in any parametrically controlled regime of the moduli space of string theory. Viable quintessence models could therefore exist only in the bulk of moduli space where numerical, even if not parametric, control could still be achieved. The systematic analysis of \cite{Cicoli:2021skd} has shown however that quintessence model building in any numerically controlled regime of the moduli space of string theory features theoretical and phenomenological challenges which are even more severe than those typical of dS constructions. This seems to suggest that single field quintessence, similarly to dS vacua, may be ruled out in string theory.

Cosmic acceleration can nevertheless arise within models containing steeper potentials, if one generalizes the prototypical single field model to include multiple real fields \cite{Brown:2017osf,Achucarro:2018vey}. The key feature that opens up this possibility turns out be the presence of non-trivial field space curvature. In \cite{Sonner:2006yn,vandeBruck:2009gp,Russo:2018akp,Cicoli:2020cfj,Cicoli:2020noz} this mechanism was studied in order to achieve viable descriptions of late-time acceleration within 4D effective models in compliance with the dS swampland conjecture. Different classes of potentials and kinetic couplings turned out to produce desirable phenomenological features and successfully fit the current cosmological data. The present phase of our universe appears there as a transient between a (past) phase of matter domination and a (future) fixed point with equation of state parameter within the range $(-1,-1/3)$.

Given the abundance of scalar fields with non-minimal coupling that generically stem from a compactification of string theory, all of this may sound very encouraging when it comes to trying to construct an honest model of quintessence with a proper stringy origin. The aim of this paper is  assessing to what extent the interesting phenomenology found in \cite{Cicoli:2020noz} can be achieved in string compactifications. To this end, we analyze typical stringy scalar potentials and try to apply the same machinery. That means searching for trajectories which start at a phase of matter domination and undergo a phase of accelerated expansion, with the cosmological parameters observed today somewhere on the trajectory. We will mainly consider two different situations. In the first case the two scalar fields are universal closed string moduli, such as e.g. axio-dilaton or the universal K\"ahler modulus in a Calabi-Yau compactification. In the other case, the scalars are instead non-universal moduli such as blow-up modes in the internal geometry. In both situations we find multifield trajectories which start from matter domination and evolve to a dark energy dominated solution. However generically none of these accelerating solutions is in agreement with current cosmological observations. The only trajectories compatible with today's cosmological observations are those with initial conditions corresponding to kinetic domination, which are hard to justify on solid grounds. 

Our analysis therefore shows that multifield quintessence, while  promising from the effective 4D point of view, turns out to be hard to realize in explicit string compactifications. This is not due to a difficulty in finding accelerating solutions, rather the problem turns out to be reproducing the observed values of the dark energy density and equation of state parameters. These findings seem to corroborate the analysis of \cite{Cicoli:2021fsd, Cicoli:2021skd} which suggests that quintessence model building in string theory is at least as challenging as the search of dS vacua.

This paper is organized as follows. In Sec. \ref{sec:2fieldmodel} we summarize the main features of the dynamics of two-field models where the underlying field space is curved, while in Sec. \ref{sec:stringy} we analyze the possibility to embed these models in string compactifications. It may be worth stressing that, in contrast to a more general class of models analyzed recently by \cite{Akrami:2018ylq}, the setup we are considering here features a conserved charge associated with shift symmetry for one of the fields. The observational differences between these models is discussed in Sec. \ref{sec:EFTs}. Besides, due to the presence of a conserved charge, semi-stable Q-balls \cite{Coleman:1985ki} could form and obstruct cosmic acceleration \cite{Kasuya:2001pr,Li:2001xaa}. Following \cite{Kasuya:2001pr}, we investigate the conditions and time-scale for the formation of Q-balls in Sec. \ref{sec:Qballs}. Finally, we implement the constraints coming from Q-ball safety within our models of interest, and in particular we apply those to the stringy trajectories found earlier. We conclude in Sec. \ref{Concl} by making further remarks and discussing possible developments of our analysis.

\section{Two-field dynamics with non-minimal coupling}
\label{sec:2fieldmodel}

The class of models which are relevant to our analysis is the class studied in \cite{Cicoli:2020noz}, describing the coupling between Einstein gravity and two real scalar fields. The underlying action is given by
\be
S[g_{\mu,\nu},\phi] \, = \, \int d^4x\, \sqrt{-g} \, \left(\frac{M_p^2}{2} \, \mathcal{R} \, - \, \frac12 \left(\partial\phi_1\right)^2 \, - \, \frac12 f(\phi_1)^2\left(\partial\phi_2\right)^2\, - \, V(\phi_1)\right) \,,
\label{eq:S}
\ee
where the crucial universal feature is the shift symmetry in the $\phi_2$ direction arising from the fact that the scalar potential is independent of the corresponding scalar field. We are interested in the study of the dynamics of these models in an FRW type background geometry
\be
ds_4^2  = - dt^2 + a(t)^2\,ds_{\mathbb{R}^3}^2  \qquad \textrm{ and } \qquad \phi_i \, = \, \phi_i(t) \ .
\ee
Note that the kinetic coupling $f(\phi_1)$, like the scalar potential $V(\phi_1)$, only depends on the field $\phi_1$. The other field $\phi_2$ is therefore a flat direction which only couples kinetically to $\phi_1$. This is typical of a complex field decomposed into polar coordinates, but also more generally for saxion and axion pairs in flux compactifications of string theory. The equations of motion are then given by
\be
\left\{
\begin{array}{ll}
\ddot{\phi}_1+3H\dot{\phi}_1-f\,f_1 \dot{\phi}_2^2+V_1=0 \,,\\[5pt]
\ddot{\phi}_2+3H\dot{\phi}_2+2\frac{f_1}{f}\dot{\phi}_2\dot{\phi}_1=0 \,,
\end{array}\right.
\label{eomphi1phi2}
\ee
where $\cdot \equiv d/dt$, $H \equiv \dot a/a$ and $f_1\equiv \partial f/\partial \phi_1$ (and similarly for $V$). For the late universe, in the presence of a barotropic fluid with pressure\footnote{We shall later assume a background of pressureless dust $\omega_b=0$ but keep it in the equations here for completeness.} $p_\gamma=\omega_b \rho_\gamma$ that evolves according to the continuity equation
\be
\dot{\rho}_b=-3 H (1+\omega_b) \rho_b\,,
\ee
the corresponding Friedman equation is
\be
3 M_p^2 H^2= \frac12\,\dot{\phi}_1^2+\frac12\,f^2 \dot{\phi}_2^2+V+\rho_b\,.
\label{eq:Fried}
\ee
It turns out to be particularly useful to introduce the following dynamical variables for the kinetic coupling
\be
k_1(\phi_1)\equiv-M_p\frac{f_1}{f} \,
\label{eq:k1Def}
\ee
and potential
\be
k_2(\phi_1)\equiv-M_p\frac{V_1}{V}\,.
\label{eq:k2Def}
\ee

The special case in which $k_1$ and $k_2$ are constant was already analyzed in \cite{Cicoli:2020cfj}, and it corresponds to a situation where both the kinetic coupling and the potential are chosen to be exponential functions of $\phi_1$, namely $f(\phi_1)=e^{-k_1\phi_1/M_p}$ and $V(\phi_1)=V_0\,e^{-k_2\phi_1/M_p}$. More general functions were the subject of further investigation in \cite{Cicoli:2020noz}. It was shown that the kinetic coupling of the massless scalar generates a novel fixed point of the dynamics, and allows for accelerating solutions even when the scalar potential is steep ($k_2\gg1$). 

Following \cite{Copeland:1997et} we define the dimensionless variables
\be
x_1\equiv \frac{\dot{\phi}_1}{\sqrt{6} H M_p}\,,\qquad x_2\equiv \frac{f \dot{\phi}_2}{\sqrt{6} H M_p}\,,\qquad y_1\equiv \frac{\sqrt{V}}{\sqrt{3} H M_p}\,,
\ee
which allow us to formulate the dynamics of the system as an autonomous system
\bea
x_1'&=& 3 x_1 (x_1^2+ x_2^2-1 ) + \sqrt{\frac{3}{2}} (-2 k_1 x_2^2 + k_2 y_1^2) - \frac{3}{2} \gamma x_1 ( x_1^2 + x_2^2 + y_1^2-1 )\,,
\label{eq:dx1} \\ [1mm]
x_2'&=&3 x_2\left(x_1^2+x_2^2-1\right) +\sqrt{6} k_1 x_1 x_2 -\frac{3}{2} \gamma  x_2 \left(x_1^2+x_2^2+y_1^2-1\right),
  \label{eq:dx2} \\ [1mm]
y_1'&=& -\sqrt{\frac32} k_2 x_1 y_1-\frac32 \gamma  y_1 \left(x_1^2+x_2^2+y_1^2-1\right)+3 y_1 \left(x_1^2+x_2^2\right),
  \label{eq:dy1}
\end{eqnarray}
where $\gamma=1+\omega_b$ and $' \equiv \frac{d}{d \ln a}$ determines the evolution in terms of e-foldings. In the general case of field dependent $k_1$ and $k_2$, these equations are to be supplemented by
\be
\phi_1'= 6\, x_1 \,.
\ee
Instantaneous fixed points are solutions compatible with $x_1'=x_2'=y_1'=0$. Note that, unless $k_1$ and $k_2$ are constant, these are not actual fixed points of the above set of equations, but are nonetheless useful for describing the underlying dynamics, provided that the energy densities of the system evolve faster than the couplings $k_1$ and $k_2$. According to the analysis in \cite{Cicoli:2020cfj}, there are six instantaneous fixed points, which we list in Tab. \ref{tab:fp} for the sake of completeness. The system is symmetric under simultaneous $\phi_1\rightarrow -\phi_1$, $k_1\rightarrow -k_1$ and $k_2\rightarrow -k_2$, 
and therefore one can focus on the $k_2>0$ half-plane without loss of generality. 

\begin{table*}[htp]
\begin{center}
\resizebox{\columnwidth}{!}{
\begin{tabular}{c c|c c c|c c |c}
\toprule
& description & $x_1$ & $x_2$ & $y_1$ & $\Omega_\phi$ & $\omega_\phi$ & stability \\
\midrule
%\hline
$\mathcal{K}_{\pm}$ & kinetic dom. & $\pm 1$ & $0$ & $0$ & 1   & 1  & unstable 
\\
$\mathcal{F}$ & fluid dom. & $0$ & $0$  & $0$ & 0   & - & unstable
\\
$\mathcal{S}$ & scaling sol.& $\frac{\sqrt{3/2}}{k_2}$ & $0$ & $\frac{\sqrt{3/2}}{k_2}$ & $\frac{3}{k_2^2}$ & $0$ & $k_2^2\ge 3$ and $k_2>2 k_1$
\\
$\mathcal{G}$ & geodesic & $\frac{k_2}{\sqrt{6}}$ & $0$  & $\sqrt{1-\frac{k_2^2}{6}}$ & $1$   & $-1+\frac{k_2^2}{3} $& $\left\{\begin{matrix}
k_2<\sqrt{3}\,,\,\,k_1>0\,,\\  k_2<\sqrt{k_1^2+6}-k_1 \\  
\end{matrix} \right.$
\\
$\mathcal{NG}$ & non-geodesic & $\frac{\sqrt{6}}{(2k_1+k_2)}$ & $\frac{\pm \sqrt{k_2^2+2k_2 k_1-6}}{2 k_1 + k_2}$ & $\sqrt{\frac{2 k_1}{2k_1+k_2}}$ & $1$ & $\frac{k_2-2k_1}{k_2+2 k_1}$ &$ \sqrt{6 + k_1^2}-k_1 < k_2 < 2k_1$\\
\bottomrule
\end{tabular}
}
\end{center}
\caption{Instantaneous fixed points of the system with a pressureless barotropic fluid $\omega_b=0$ and two scalar fields for $k_2>0$. The stability region is given in the last column.}
\label{tab:fp}
\end{table*}

In order to test the model against late time observations, it may be  enlightening to directly study the time evolution of the relevant late time observables. We denote the equation of state parameter for the quintessence field  by $\omega_\phi$, and the corresponding density parameter by $\Omega_\phi$. Re-expressing them in terms of our dynamical variables $\left(x_1,x_2,y_1\right)$, we find
\be
\omega_\phi\,=\,\frac{p_\phi}{\rho_\phi}\,=\,\frac{x_1^2+x_2^2-y_1^2}{x_1^2+x_2^2+y_1^2}\,,
\label{eq:omega}
\ee
and
\be
\Omega_\phi\,=\,x_1^2+x_2^2+y_1^2 \,.
\label{eq:Omega}
\ee
The evolution of these quantities obeys the differential equations
\be\begin{split}
\Omega_\phi'&=-3 \left(\Omega_\phi -1\right) \Omega_\phi  (\omega_b -\omega_\phi)\,,  \\
\omega_\phi'&=(\omega_\phi -1) \left(-k_2 \sqrt{3(\omega_\phi +1)\Omega_\phi -6 x_2^2}+3(1+ \omega_\phi) \right).
\end{split}
\ee
It is immediately clear from Tab. \ref{tab:fp} that the fixed points cannot describe our current universe with the cosmological observables $\Omega_\phi^{(0)}\sim0.7$ and $\omega_\phi^{(0)}\sim-1$. We should instead look for transients in the $(x_1, x_2, y_1)$ parameter space which start from matter domination, exhibit a (short) period of accelerated expansion\footnote{
The number of e-foldings from the end of matter domination until today may be roughly estimated to be $\Delta N = \ln\left(\nicefrac{a_0}{a}\right) \sim \frac13\,\ln\left(\nicefrac{\Omega_\Lambda^{(0)}}{\Omega_{\textrm{m}}^{(0)}}\right) \sim 0.28$.} before passing through our current universe. The universal regime of matter domination occurs when the barotropic fluid is the dominant contribution in the energy-momentum tensor. This implies the necessity of choosing initial conditions close to the origin $(0,0,0)$, where both $\omega_\phi$ and $\Omega_\phi$ are small.
 
Summarizing, in \cite{Cicoli:2020noz} various different choices of functional dependence for the scalar potential and the kinetic coupling were investigated. Generically it was observed that accessing the stability region for the non-geodesic fixed point produces trajectories which display cosmic acceleration in the transient path towards the fixed point itself. This was found to hold both in the case where $k_1$ and $k_2$ are constant and when they vary with $\phi_1$, without substantial difference at a qualitative level. However empirically, an $\mathcal{O}(10)$ hierarchy between $k_1$ and $k_2$, i.e. $k_1 \gtrsim \mathcal{O}(10)\, k_2$, seemed to be needed in order to obtain late time observables compatible with their current experimental range. This particular feature might already intuitively generate the suspicion that a possible compatibility of these constructions with a proper stringy embedding may be contrived.

As we will see later on, it turns out to be difficult to shoot from the origin and have full control over all the possible transients. An alternative approach could be to require a transient which is compatible with current observations in the first place, and integrate backwards to see the past trajectory and associated initial conditions. It may be worth stressing that, although we only have two observables to constrain the initial value of three coordinates, eqs. \eqref{eq:omega} and \eqref{eq:Omega} fully determine the observationally viable region to be around $\left(0,0,\sqrt{\Omega_\phi^{(0)}}\right)$.

\section{Stringy two-field quintessence}
\label{sec:stringy}

The goal of this section is to investigate concrete string inspired two-field models and assess the existence of viable accelerating trajectories in agreement with today's observations. As we have already anticipated earlier, the possibility of producing a hierarchy between $k_1$ and $k_2$ appears rather unnatural within a stringy context. Indeed, we will show the absence of quintessence models compatible with the data, both within a class of examples where the quintessence fields are universal moduli, and within a different class in which the same role is played by blow-up modes. Note that this result crucially relies on the assumption of matter-dominated initial conditions $\Omega_\phi\sim0$. In contrast, we will also show that good fits for current observations are in principle possible if such an assumption is dropped. 
%While the asymptotic past of these trajectories lies in the kinetic dominated fixed points, we argue that there is generically a point in the past which could be interpreted as matter dominated.

\subsection*{Universal closed string moduli}

Effective Lagrangians of the form \eqref{eq:S} can arise in the context of string compactifications. Consider for example a situation where all but one closed string moduli are stabilized at specific vacuum expectation values by means of suitable internal fluxes and sources. This leaves one single complex field, whose real and imaginary parts are then to be mapped into our real fields $(\phi_1,\phi_2)$.

A particularly thorough construction of this type was presented in \cite{Saltman:2004sn}, where type IIB compactifications on Calabi-Yau 3-folds were considered, with the addition of non-vanishing internal R-R and NS-NS 3-form fluxes. Flux quantization was properly discussed, and tadpole cancellation was found to require spacetime filling O3 planes and parallel D3-branes. In this context, the authors were able to show that, for some explicit choices of the flux quanta, it was actually possible to stabilize all complex structure moduli and the axio-dilaton at a non-supersymmetric locus. This generates a runaway for the overall K\"ahler modulus $T= {\rm Vol}(\Sigma_4) + {\rm i} \int_{\Sigma_4} C_{(4)}$ where ${\rm Vol}(\Sigma_4)$ is the volume of the internal overall 4-cycle $\Sigma_4$ in string units, while $C_{(4)}$ is the R-R 4-form. Though the explicit solutions that they provide are for the special choice of Calabi-Yau given by a $\mathbb{Z}_2$ orbifold of the six-torus, similar examples exist for Calabi-Yau 3-folds \cite{Gallego:2017dvd}. The constructions involve $0$, $2$ or $4$ D3 branes, a positive vacuum energy and all the modes but $T$ have been stabilized. 

The effective 4D description turns out to be given by $\mathcal{N}=1$ supergravity coupled to chiral multiplets. The corresponding flux-induced superpotential is independent of the overall volume modulus, and hence a no-scale structure emerges. Once all the chiral fields but the volume $T$ are stabilized, the effective description only involves one single complex field, for which the kinetic Lagrangian is fully specified by the following K\"ahler potential
\be
K =-3 \ln \left(T+\overline{T}\right) ,
\ee
while the effective scalar potential reads 
\be
V_{\rm eff} = \frac{V_0}{\left(T+\overline{T}\right)^3}\,M_p^4\,,
\ee
where $V_0>0$ appears after evaluating all the other scalar fields in their constant vacuum expectation values. After choosing the convenient parametrization\footnote{The parametrization introduced here yields a canonically normalized $\phi_1$, while the kinetic term for $\phi_2$ couples exponentially to $\phi_1$.} $T\,=\,e^{\sqrt{\frac23}\phi_1/M_p}\,+\, {\rm i} \, \frac{\phi_2}{M_p}$, one precisely recovers exponential potential and kinetic coupling, with 
\be
k_1= \sqrt{\frac23} \qquad\text{and}\qquad k_2=\sqrt{6} \,. 
\ee
While in \cite{Saltman:2004sn, Gallego:2017dvd} these compactifications with run-away volume modulus were considered with the aim of stabilizing the remaining directions by adding higher derivative or non-perturbative effects, we want to use them to construct two-field quintessence configurations instead.

Beyond the above stringy type IIB model, one might think of considering a whole class of string inspired models of the same type, by simply considering the following K\"ahler potential for a chiral superfield $X$
\be
K =-p \ln \left(X+\overline{X}\right) \,+\, \dots\,,
\label{Kaehler_p}
\ee
where $X$ is no longer assumed to be identified with the volume modulus, but is instead allowed to be any complex field appearing in the compactification at hand, and which has not yet been stabilized. The dots in \eqref{Kaehler_p} denote $X$ independent contributions. This type of K\"ahler potential is quite common for closed string moduli. Should in addition the superpotential $W$ be independent\footnote{Note that this is an immediate generalization of what happens in the no-scale supergravity model arising in type IIB Calabi-Yau orientifolds, as shown in \cite{Ferrara:2002bt,Burgess:2020qsc}.} of $X$, $\partial_X W=0$, the supergravity F-term potential
\be
V_F= e^K \left(K^{X\overline{X}} D_X W D_{\overline{X}} \overline{W} - 3 |W|^2 \,+\, \dots\right) M_p^4 
\ee
would then turn out to depend on $X$ only through the $e^K$ factor, and hence we have
\be
V_F\propto\left(X+\overline{X}\right)^{-p}\,.
\label{VF_propto_X}
\ee
Note that the above class contains the type IIB case of \cite{Saltman:2004sn} discussed above, for which $p=3$. Another prototypical situation within this class is given by heterotic compactifications on $\mathrm{SU}(3)$ structure manifolds. The associated 4D effective description is again a minimal supergravity coupled to chiral multiplets, but this time the role of the field $X$ not appearing in the superpotential is given by $S=e^{-\varphi}+{\rm i} \,a$ where $\varphi$ is the 4D dilaton while its axionic counterpart $a$ is the 4D dual of $B_{(2)}$, $\star da=dB_{(2)}$. Also in this case one could imagine stabilizing all of the complex structure and  K\"ahler moduli, leaving the $S$ direction as a runaway (see \cite{Cicoli:2013rwa} for a systematic analysis of heterotic moduli stabilization). This would then, by construction, fit our discussion when $p=1$. 

The case with $p=2$ can be reproduced instead by considering type IIB compactified on orientifolds of CY 3-folds which are K3 or $T^4$ fibrations over a $\mathbb{P}^1$ base. In this situation the CY volume in string units reads $\mathcal{V}={\rm Vol} (K3) {\rm Vol} (\mathbb{P}^1) = \sqrt{\tau_1} \tau_2$ where ${\rm Vol} (K3) = \tau_1$ and ${\rm Vol} (\mathbb{P}^1) = \tau_2/\sqrt{\tau_1}$ \cite{Cicoli:2011it,Cicoli:2016xae,Cicoli:2017axo}. Clearly $K=-2\ln\mathcal{V}$ reproduces the $p=2$ case if $X$ is identified with $T_2= \tau_2 + {\rm i}\int_{\Sigma_4^{(2)}} C_{(4)}$ and $T_1$, together with all the other geometric moduli, is stabilized at a non-supersymmetric locus. Finally, for other intermediate values of $p$, one could imagine starting from the model with seven chiral multiplets arising from compactification of M-theory on a $\mathbb{Z}_2^3$ orbifold of the seven-torus, and start identifying some of the fields among them. The extreme $p=7$ case is obtained by identifying all of them in a single field $Z={\rm Vol}(\Sigma_3) + {\rm i} \int_{\Sigma_3} A_{(3)}$ where ${\rm Vol}(\Sigma_3)$ is the volume of an internal 3-cycle $\Sigma_3$ in 11D units, while $A_{(3)}$ is the M-theory 3-form. A panoramic view of relevant stringy examples yielding models within this class is offered in Tab. \ref{tab:stringy_p}.

\begin{table}[htp]
\begin{center}
\begin{tabular}{l c c c c c}\toprule

$p$ & $X$ & Theory & Sources &  $\mathcal{M}_{\rm internal}$ & References \\
\midrule
$1$ & $S = e^{-\varphi} + {\rm i}\, a$ & Heterotic & --- & $\mathrm{SU}(3)$ str. & \cite{Font:1990nt} 
\\
$2$ & $T_2 = {\rm Vol}(\Sigma_4^{(2)}) + {\rm i} \int_{\Sigma_4^{(2)}} C_{(4)}$ & Type IIB & D3/D7, O3/O7 & K3-fibered $\mathrm{CY}_3$  & \cite{Cicoli:2011it,Cicoli:2016xae,Cicoli:2017axo} 
\\
$3$ & $T = {\rm Vol}(\Sigma_4) + {\rm i} \int_{\Sigma_4} C_{(4)}$ & Type IIB & D3/O3 & $\mathrm{CY}_3$  & \cite{Saltman:2004sn} 
\\
$7$ & $Z = {\rm Vol}(\Sigma_3) +{\rm i}\int_{\Sigma_3} A_{(3)}$ & M-theory & KK6/KKO6 & $\mathrm{G}_2$ str. & \cite{Blaback:2019zig}
\\
\bottomrule
\end{tabular}
\end{center}
\caption{Examples of string compactifications yielding 4D two-field models with exponential kinetic coupling and exponential runaway potential. The two-field models arise after fixing the rest of the moduli present in the given setup.}
\label{tab:stringy_p}
\end{table}

\begin{figure}[h!]
	\centering
	\begin{minipage}[b]{0.45\linewidth}
	\centering
	\includegraphics[width=\textwidth]{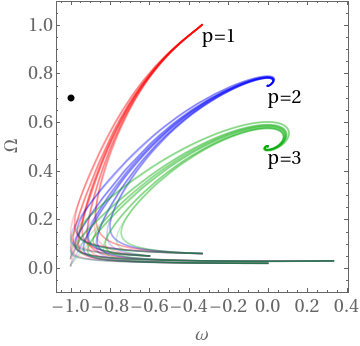}
    \end{minipage}
	\hspace{0.5cm}
	\begin{minipage}[b]{0.435\linewidth}
	\centering
	\includegraphics[width=\textwidth]{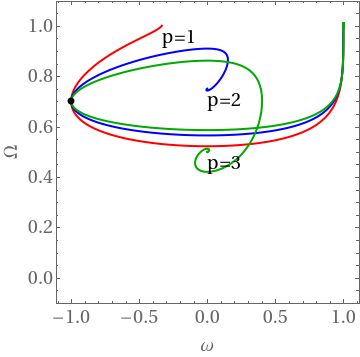}
	\end{minipage}
\caption{Dynamical trajectories in the $(\omega,\Omega)$ plane for stringy models with $k_1$ and $k_s$ specified by \eqref{k1k2fromp}. {\it Left:} Initial condition are chosen to be matter domination. None of the trajectories go through the black dot representing the correct current value for late time observables. {\it Right:} The trajectories are required to pass through the observed current values of $(\omega,\Omega)$.}
\label{fig:oO_universal}
\end{figure}

Note that $V_F$ as illustrated in \eqref{VF_propto_X} only depends on the real part of $X$, leaving the imaginary part as a flat direction. This is in perfect agreement with our construction of the two-field models described above, where the kinetically coupled field has a flat potential.
The kinetic part of the action reads
\be
\mathcal{L}_{\textrm{kin}} = - \sqrt{-g}\,K_{X\overline{X}}\, \partial X \partial \overline{X}\, M_p^2 
= - \frac12\sqrt{-g}\,\left((\partial \phi_1)^2+\frac{p}{2}\ e^{2 \sqrt{2/p}\ \phi_1/M_p}(\partial \phi_2)^2\right) ,
\ee
where
\be
\frac{\phi_1}{M_p}\equiv \,\sqrt{\frac{p}{2}} \ln \left(\frac{X+\overline{X}}{2}\right)  
\qquad\text{and}\qquad \frac{\phi_2}{M_p}\equiv  \frac{X-\overline{X}}{2\,{\rm i}} \,.
\ee
One can therefore define the kinetic coupling function as
\be
f(\phi_1)=\sqrt{\frac{p}{2}}\,e^{\sqrt{2/p}\ \phi_1/M_p}\,.
\ee
The potential depends on $\phi_1$ as
\be
V(\phi_1) = \frac{V_0}{\left(X+\overline{X} \right)^p}= \frac{V_0}{2^p}\, e^{-\sqrt{2 p}\ \phi_1/M_p}\,. 
\ee
So we see that in these simple supergravity models $k_1$ and $k_2$ are constant and their values are expressed in terms of $p$ as
\be
k_1= \sqrt{2/p} \qquad\text{and}\qquad k_2=\sqrt{2p} \,. 
\label{k1k2fromp}
\ee
Let us now look at the qualitative and quantitative features of dynamical trajectories. At first glance, from \eqref{k1k2fromp} we can already see that $k_2/k_1=p$, which rules out the desired hierarchy that made it possible in \cite{Cicoli:2020noz} to achieve phenomenologically viable transients. This confirms our initial suspicions, at least for integer values of $p$, which are those admitting a clear stringy interpretation. Indeed, if we first start by giving initial conditions near matter domination, we find the trajectories displayed in the left panel of Fig. \ref{fig:oO_universal}. As is evident from there, though $p=1$ admits an accelerating regime when $\Omega_\phi=0.7$, neither of the $p=1,2,3$ cases allows for a viable description of the present day universe. This is due to the fact that in regimes close to $\Omega_\phi=0.7$, the equation of state parameter deviates too much from $-1$. If we considered higher values of $p$, the associated curves would bend even more sharply down to the right, only making the situation worse.
%thus further increasing the value of $\omega_\phi$, in particular above the threshold value of $-1/3$, which is where cosmic acceleration ceases to be possible.

In line with what anticipated above, in order to have a viable transient through the observationally preferred region of the $(\omega,\Omega)$ plane, one would need to have $p\le 0.2$, which rules out the possibility of $\mathcal{NG}$ dynamics playing a role in this class of models. Needless to say, such small and non-integer values for $p$ are difficult to motivate from a UV point of view. Note that this upper bound on $p$ corresponds to $k_2<0.6$ as found in \cite{Agrawal:2018own}.

However, if we instead look for trajectories that pass through observationally viable points, we find that these do exist. They may be found by using the correct current values of late time observables as conditions to integrate the equations \eqref{eq:dx1}, \eqref{eq:dx2} and \eqref{eq:dy1}. By repeating the analysis for $p=1,2,3$, we see that the future fixed points differ but the behavior in the past is qualitatively similar for every $p$. This situation is shown in the right panel of Fig. \ref{fig:oO_universal}.

\begin{figure}[h!]
	\centering
	\begin{minipage}[b]{0.45\linewidth}
	\centering
	\includegraphics[width=\textwidth]{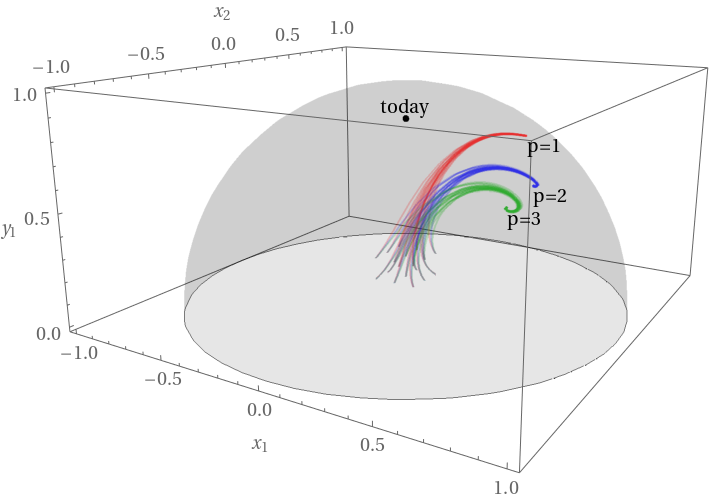}
    \end{minipage}
	\hspace{0.7cm}
	\begin{minipage}[b]{0.45\linewidth}
	\centering
	\includegraphics[width=\textwidth]{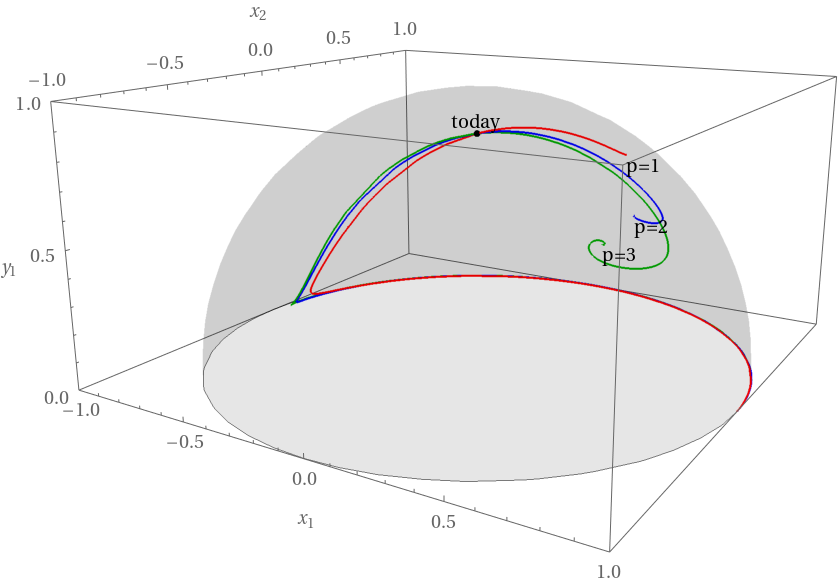}
	\end{minipage}
\caption{Dynamical trajectories in the $(x_1,x_2,y_1)$ space for stringy models with $k_1$ and $k_s$ specified by \eqref{k1k2fromp}. {\it Left:} Initial condition are chosen to be matter domination. All trajectories start from the center of the hemisphere and approach the surface without going through the black dot. {\it Right:} All trajectories pass through the correct today's values of late time observables. In this case all trajectories originate from kinetic domination initial conditions.}
\label{fig:xxy_universal}
\end{figure}

In Fig. \ref{fig:xxy_universal}, we display the same solutions but in the $(x_1,x_2,y_1)$ space, for matter domination initial conditions (left panel) and for trajectories which feature transients compatible with today's observations (right panel). All the trajectories in the right plot feature kinetic domination initial conditions since they move towards the observable point from the kinetic dominated regime $\mathcal{K}_-$, and universally originate from $\mathcal{K_+}$ in the asymptotic past, as long as $x_2^{\rm today}\neq 0$. While these trajectories do not originate from matter domination in the sense of fluid domination, there are points along the past trajectory which could be interpreted as matter dominated. In particular, the points on the trajectories where $\omega_\phi=0$ could be indistinguishable from pure matter cosmology since the quintessence field mimics the barotropic fluid with $\omega_b=0$. In this sense, one might say that in this cosmological phase the universe would effectively behave as a matter dominated universe, where $\phi$ counts as part of (actually roughly half of) the matter content of the universe. While these phases are generically present, it is hard to motivate them as initial conditions in comparison with the more standard case of fluid domination initial conditions.

\subsection*{Non-universal moduli: blow-up modes}

In the context of Calabi-Yau compactifications, particularly simple examples that one can work with are toroidal orbifolds. These geometries present singularities at the location of the fixed points of the corresponding orbifold action. This issue is overcome via the blow-up procedure which provides an explicit prescription for resolving these singularities into compact and smooth manifolds, at the price of introducing a number $N$ of non-universal moduli controlling the size of these blow-up modes. Effective 4D models for blow-up modes can be obtained from a class of type IIB orientifold flux compactifications with D3/O3 as well as D7/O7 spacetime filling sources. The corresponding internal manifold is a Calabi-Yau 3-fold of a so-called `weak Swiss-cheese' form \cite{Cicoli:2014sva,Cicoli:2018tcq}. In this case, the total internal volume may be written as
\be
\mathcal{V} \, = \, f_{3/2}(\tau_j) \,-\, \sum\limits_{i=1}^{N}\lambda_i \tau_i^{3/2} \,,
\ee
where $\left\{\tau_j\right\}$ are the universal moduli controlling the volume, $f_{3/2}$ is a homogeneous function of degree $3/2$, while $\left\{\tau_i\right\}$ are blow-up moduli controlling the volume of `diagonal del Pezzo' divisors \cite{Cicoli:2011it}. Let us focus for simplicity on the simplest case with just one universal modulus $\tau_b$ and one blow-up mode $\tau_s$ with $\tau_s\ll\tau_b$. This hierarchy can be achieved if at leading order $\tau_b$ is stabilized exponentially large by balancing $\alpha'$ against loop corrections as in \cite{Antoniadis:2018hqy, Burgess:2022nbx}. The non-universal modulus $\tau_s$ can then be fixed at subleading order by the interplay between higher order perturbative corrections as in \cite{Cicoli:2016chb}. At this level of approximation, the axionic partner of $\tau_s$ is still massless due to its perturbative shift symmetry. This situation therefore would be described by the Lagrangian (\ref{eq:S}) where $\phi_1$ is the canonically normalized blow-up mode and $\phi_2$ the corresponding axion. 

In the $\tau_s\ll\tau_b$ limit, the effective K\"ahler potential for $\tau_s$ is  power-law rather than logarithmic 
\be
K = -2\ln \mathcal{V}\simeq -3 \ln \tau_b +2\left(\frac{\tau_s}{\tau_b}\right)^{3/2}\,,
\ee
implying the following form of the kinetic terms (for $\tau_s= (T_s+\overline{T}_s)/2$)
\be
\mathcal{L}_{\textrm{kin}} = - \sqrt{-g}\,K_{T_s\overline{T}_s}\, \partial T_s \partial \overline{T}_s\, M_p^2 
= - \frac12\sqrt{-g}\,\left((\partial \phi_1)^2+\left(\frac{M_p}{\phi_1}\right)^{2/3}(\partial \phi_2)^2\right) ,
\ee
where
\be
\frac{\phi_1}{M_p}\equiv \,\frac{2}{\sqrt{3}} \left(\frac{\tau_s}{\tau_b}\right)^{3/4}  
\qquad\text{and}\qquad \frac{\phi_2}{M_p}\equiv  \left(\frac34\right)^{1/3}\frac{(T_s-\overline{T}_s)}{2\,{\rm i}\,\tau_b} \,.
\label{canNorm}
\ee
Thus the kinetic coupling function is given by
\be
f(\phi_1)=\left(\frac{M_p}{\phi_1}\right)^{1/3}\,.
\ee
The scalar potential generated by perturbative corrections (string loops or higher derivative effects) has been studied in concrete Calabi-Yau orientifold examples with explicit brane setup and tadpole cancellation \cite{Cicoli:2011qg,Cicoli:2016xae,Cicoli:2017axo} and can take slightly different forms. Its behavior is always power-law
\be
V(\phi_1)=V_0 \left(\frac{M_p}{\phi_1}\right)^{\pm 2/3}\qquad\text{or}\qquad V (\phi_1) = \frac{V_0}{C-\left(\phi_1/M_p\right)^{2/3}}
\ee
where $C$ is a positive $\phi_1$-independent quantity which in explicit Calabi-Yau models turns out to be of $\mathcal{O}(1)$. These models motivate the following more general class of choices for $f$ and $V$ 
\be
f (\phi_1) =  \left(\frac{M_p}{\phi_1}\right)^{p_1}  \qquad \text{and} \qquad V (\phi_1)\, =  \left(\frac{M_p}{\phi_1}\right)^{p_2}
\frac{V_0}{C \left(M_p/\phi_1\right)^{p_2}-1} \,,
\ee
with $p_1>0$, $p_2>0$ for $C\neq 0$, and $p_2$ which can have either sign for $C=0$. This leads to
\be
\frac{k_2}{k_1} = \left(\frac{p_2}{p_1}\right)\frac{1}{1-C \left(M_p/\phi_1\right)^{p_2}}\,.
\label{ratio}
\ee
If $C=0$, accelerating multifield solutions with observationally viable transients could in principle exist if $p_2>0$ but they would require $k_2/k_1\lesssim 0.1$.\footnote{For $p_2<0$ the ratio $k_2/k_1$ is negative, implying that no non-geodesic dynamics can be present according to the analysis of \cite{Cicoli:2020noz}.} This regime could be achieved for $p_2 \ll p_1$ but explicit stringy examples do not seem to reproduce this hierarchy. In fact, the concrete Calabi-Yau models of \cite{Cicoli:2011qg,Cicoli:2016xae,Cicoli:2017axo} would lead to $k_2/k_1=2$. On the other hand, if $C\neq 0$, the term proportional to $C$ in (\ref{ratio}) would dominate since $\phi_1\ll M_p$ in the $\tau_s\ll \tau_b$ limit, as can be seen from (\ref{canNorm}). In this case (\ref{ratio}) would however be negative for $p_2>0$, leading to the absence of any non-geodesic dynamics as shown in \cite{Cicoli:2020noz}. Notice that, even allowing $p_2<0$, the system would still not feature any phenomenologically viable transient since $k_2/k_1$ would become:
\be
\frac{k_2}{k_1} \simeq  \left(\frac{|p_2|}{C p_1}\right)\left(\frac{M_p}{\phi_1}\right)^{|p_2|}\gg 1\qquad\text{for}\qquad \phi_1\ll M_p\,,
\ee
unless $|p_2|\ll 1$ which is however very unnatural from the UV point of view.

We have therefore shown that blow-up modes enrich the class of multifield models by including power-law potentials and kinetic couplings together with a field-dependent $k_2/k_1$ ratio. Unfortunately, they do not lead to qualitatively new results regarding the possibility to achieve a non-geodesic multifield dynamics which reproduces the observational features of today's universe.

\section{Conserved currents and observables in two-field quintessence}
\label{sec:EFTs}

In this section we clarify the relation between two-field quintessence models that feature a conserved current like the one under consideration \cite{Cicoli:2018ccr}, and those that do not \cite{Akrami:2020zfz}. In order to do so we will try to reduce the problem to a single field one. In the former case this can be done by exploiting the conserved charge to eliminate the light degree of freedom from the action, thereby obtaining an action with a time-dependent effective potential. In the latter case, whenever there is a mass hierarchy (be it at the background or perturbation level) one can integrate out the heavy degree of freedom to find a low energy effective action. This procedure allows us to show that, though the two classes of models descend from a similar Lagrangian, they are in fact observationally distinguishable. 

The starting point for both models is an action which generalizes (\ref{eq:S}) by allowing the potential to depend on both fields, i.e. $V=V(\phi_1,\phi_2)$. The first equation of motion in (\ref{eomphi1phi2}) is unaltered, while the second acquires an additional contribution proportional to the derivative of $V$ with respect to $\phi_2$
\be
\ddot{\phi}_2+3 H \dot{\phi}_2+ 2\frac{f_1}{f} \dot{\phi_2}\dot{\phi_1}=-\frac{V_2}{f^2}\,.
%\label{eq:eomphi1}
\ee
When $V_2$ vanishes identically, i.e. for our original model \eqref{eq:S}, there is a conserved current and and associated conserved charge
\be
J^\mu=\sqrt{-g} f^2 \partial^\mu \phi_2\qquad\text{and}\qquad Q=\int d^3x J^0\,.
\ee
Using the charge density $J^0=a^3 f^2 \dot{\phi_2}\equiv q$ one can eliminate $\dot{\phi_2}$ from the first equation in (\ref{eomphi1phi2}), thereby obtaining the effective equation of motion
\be
\ddot{\phi}_1+3 H \dot{\phi}_1-\frac{f_1 q^2}{f^3 a^6}+ V_1=0\,,
\label{eq:eomphi1Eff}
\ee
which prompts the definition of a time-dependent `effective potential'
\be
V_{\rm eff}(\phi_1,a)=V(\phi_1)+\frac{q^2}{2 f^2 a^6}\,.
\ee
The main point is that, regardless of the time dependence of $V_{\rm eff}$, eq. \eqref{eq:eomphi1Eff} can be obtained from a two derivative Lagrangian, and therefore perturbations in such a background propagate at the speed of light. In other words the speed of sound is unity: $c_s=1$.

Let us now turn our attention to the $V_2\neq 0$ case analyzed in \cite{Akrami:2020zfz}. We define $m_1^2\equiv V_{11}$ and assume that $m_1\gg H$ such that $\phi_1$ is a heavy field, and can therefore be integrated out. We are interested in slow-roll solutions to this system ($\ddot{\phi}_1\approx 0$ and $\ddot{\phi}_2\approx 0$) with negligible velocity for the heavy field $\dot{\phi}_1\approx 0$. In this regime, the equation of motion for the heavy scalar reads
\be
f\,f_1\,\dot{\phi}_2^2\approx V_1 \,.
\label{eq:eomphi1}
\ee
Solving this equation allows us to integrate out the heavy field, thereby obtaining an effective action for the light scalar driving quintessence. As we now show following \cite{Pedro:2019klo}, this action involves higher derivative terms of the form $(\dot{\phi_2})^{2 n}$. Let us define
\be
\phi_1\equiv \langle\phi_1\rangle (1+\delta)\,,
\label{eq:phiH}
\ee
where $\langle\phi_1\rangle$ is the solution to $V_1=0$ when $\phi_1$ relaxes to its minimum after the end of the quintessence epoch. During dark energy domination, the field $\phi_2$ is displaced from its minimum, and so the solution to $V_1=0$ is shifted to $\phi_1 = \langle\phi_1\rangle (1+c_0)$ with $c_0 \ll 1$ since the mass of $\phi_1$ is much larger than $H$. Moreover (\ref{eq:eomphi1}) implies that $\phi_1$ does not sit at the effective minimum of the potential during quintessence since the velocity of $\phi_2$, even if tiny, is non-zero. Hence throughout the whole dynamics $\delta\ll 1$ and can be expanded in powers of $\dot{\phi}_2^2$ as
\be
\delta=\sum_{n\ge 0} c_{2n}\left(\frac{\dot{\phi}_2^2}{\Lambda^4}\right)^n \,.
\label{eq:delta}
\ee
One can then solve eq. \eqref{eq:eomphi1} order-by-order in the velocity of the light field. Expanding in powers of $\dot{\phi}_2$ yields
\be
\begin{split}
\bar{f} \bar{f}_1 \dot{\phi}_2^2 &+ c_2 \langle\phi_1\rangle \left(\bar{f}_1^2 + \bar{f} \bar{f}_{11} \right)\frac{\dot{\phi}_2^4}{\Lambda^4}+\dots
\\[3pt]
&=\, \bar{V}_1 + c_2 \langle\phi_1\rangle \bar{V}_{11}\frac{\dot{\phi}_2^2}{\Lambda^4} + \langle\phi_1\rangle\left(c_4 \bar{V}_{11} + \frac{c_2^2}{2}\langle\phi_1\rangle \bar{V}_{111} \right)\frac{\dot{\phi}_2^4}{\Lambda^8} +\dots\,,
\end{split}
\label{eq:aux}
\ee
where the dots on both sides stand for neglected $\mathcal{O}\left({\dot{\phi}_2}^6\right)$ terms, and we defined the notation $\bar{f}\equiv f(\langle\phi_1\rangle(1+c_0))$ and $\bar{V}\equiv V(\langle\phi_1\rangle(1+c_0),\phi_2)$, and similarly for the corresponding derivatives with respect to $\phi_1$. From eq. \eqref{eq:aux} one finds the expressions for the $c_{2n}$ coefficients. The first three terms take the form
\be\begin{split}
&V_1\left(\langle\phi_1\rangle(1+c_0),\phi_2\right)=0 \,, \\
&c_2=\frac{\bar{f} \bar{f}_1 \Lambda^4}{\langle\phi_1\rangle \bar{V}_{11} } \,,  \\
&c_4= \frac{2 \left(\bar{f}_1^2  + \bar{f} \bar{f}_{11}\right) \bar{V}_{11} - \bar{f} \bar{f}_1 \bar{V}_{111} }
{2 \langle\phi_1\rangle \bar{V}_{11}^3}\,\bar{f} \bar{f}_1 \Lambda^8\,.
\label{eq:cnSol}
\end{split}
\ee
The first equation in \eqref{eq:cnSol} implicitly defines $c_0$ and implies that, to leading order, the heavy field adiabatically follows its $\phi_2$-dependent minimum. One can use the solution to the heavy field equation of motion, eqs. \eqref{eq:phiH}, \eqref{eq:delta} and \eqref{eq:cnSol}, to integrate it out and find the effective field theory for the light scalar degree of freedom at the level of the background.

By ignoring the contribution of the heavy field kinetic term, one can write the single-field effective action as 
\be
\begin{split}
\mathcal{L}_{\rm eff}=\sqrt{-g}\Bigg( &\frac12 \bar{f}^2 \dot{\phi}_2^2 -\bar{V}(\phi_2)+ \frac12 \frac{(\bar{f} \bar{f}_1)^2}{\bar{V}_{11}}\dot{\phi}_2^4
\\
&+\frac16\frac{(\bar{f} \bar{f}_1)^2}{\bar{V}_{11}^3} \left[3 \left(\bar{f}_1^2  + \bar{f} \bar{f}_{11} \right) \bar{V}_{11} - \bar{f} \bar{f}_1 \bar{V}_{111}\right]  \dot\phi_2^6 + \mathcal{O}(\dot{\phi}_2^8)\Bigg).
\end{split}
\label{eq:PofX1}
\ee
Thus the low energy action of the light scalar $\phi_2$ is of the $P(X,\phi_2)$ form where $X\equiv \frac12 \dot{\phi}_2^2$. Interactions between the heavy and the light scalar generically give rise not only to corrections to the scalar potential and kinetic terms of the light field, but also to higher derivative terms. It is precisely these that play a crucial role in the observational signatures of this model. Written in this manner, it is clear that the model of \cite{Akrami:2020zfz} falls into the class of K-essence models whose most salient feature is the reduction of the speed of sound for the scalar perturbations induced by the presence of higher derivative terms. In fact, the speed of sound can be written as
\be
c_s^{-2}=1+2 X \frac{P_{XX}}{P_X} = 1\,+\, 4 \frac{ \bar{f}_1^2}{\bar{V}_{11}} \dot{\phi}_2^2\,+\,4 \frac{\bar{f}_1^2\left((\bar{f}_1^2  + 
3 \bar{f} \bar{f}_{11}) \bar{V}_{11}- \bar{f} \bar{f}_1 \bar{V}_{111} \right)}{\bar{V}_{11}^3}\dot{\phi}_2^4\,+\,\mathcal{O}(\dot{\phi}_2^6)\,.
\label{eq:csEFTb}
\ee
Taking the explicit model of \cite{Akrami:2018ylq}, where $\phi_1=r$, $\phi_2=\theta$, $f(\phi_1)=\phi_1/M_p$ and $V=V_0 -\alpha \phi_2+\frac12 m^2(\phi_1-\langle\phi_1\rangle)^2$ one finds \cite{Achucarro:2015rfa,Pedro:2019klo}
\be
c_s^{-2}=1+ \left(\frac{2\dot{\phi}_2}{m M_p}\right)^2+\frac14 \left(\frac{2\dot{\phi}_2}{m M_p}\right)^4 +\, \dots \,,
\ee
in perfect agreement with the results of \cite{Akrami:2018ylq}.

\section{Q-ball formation in two-field models with a conserved charge}
\label{sec:Qballs}

In this paper we have investigated the possible existence of stringy two-field quintessence models compatible with late-time cosmological observations. Since our models enjoys a shift symmetry in the $\phi_2$-direction, they might develop instabilities at a non-perturbative level associated with spontaneous formation of solitonic objects that are charged under this symmetry. Solitonic, non-topological field configurations of a complex scalar field with $U(1)$ symmetry are known as Q-balls. These extended objects are the minimum energy configuration for a given charge. They are completely stable by virtue of charge conservation. Producing Q-balls during quintessence effectively screens the dark energy, abruptly stopping the accelerating dynamics. In order to make our analysis of two-field models more complete, we want to conclude by studying the constraints imposed by stability against spontaneous Q-ball formation.

Intuitively, Q-balls exist if the interacting potential is less than the free field part ${V(\phi)<m^2|\phi|^2}$. Then a compact, self-sustaining configuration is energetically favored over a cloud of massive, weakly interacting particles. Quantitatively, \cite{Coleman:1985ki} showed that the condition for the formation of these objects is a minimum of the function $V(\phi)/|\phi|^2$ at non-zero values of $\phi\neq 0$. However, the derivation of this condition in \cite{Coleman:1985ki} assumes the existence of a true vacuum at $\phi=0$ which is not the case for our quintessence models. Moreover, in the situation of our interest, the complex scalar field theory is coupled to gravity in an FRW background.

Due to the absence of a true vacuum, one might say that rather than Q-balls, we have what \cite{Krippendorf:2018tei} called PQ-balls. In particular, a spherical Q-ball solution for exponential potentials and kinetic couplings has divergent total charge and energy with densities that have a $1/r^2$ decay at infinity. Since the total charge and energy is infinite, the configurations cannot minimize the energy for fixed charge. In the quintessence case, the background is not a true vacuum, but rather an evolving state with finite and constant charge and energy density, and so the total background charge and energy is already necessarily divergent. As a consequence, while Q-balls may not be totally stable in this background, we argue that there are no fundamental obstructions to localized objects which behave like Q-balls.\footnote{This situation is similar to the one occurring with field theory domain walls coupled to gravity. Within the gravitational theory, non-perturbative processes may create expanding Coleman-De Luccia bubbles \cite{Coleman:1980aw} with a non-zero probability. Local Q-ball formation in a gravitating background may be seen as a generalization of these bubbles that includes the extra $U(1)$ charge.}

We are now interested in the formation of these objects during the evolution of the system. An abundance of Q-balls formed anywhere along the trajectory 
would not only be observationally problematic, but it would even drastically change the equation of state for the quintessence field to that of matter or radiation (depending on the stability mentioned above), as the Q-balls screen the expanding effects of the classical background solution. 

Following \cite{Kasuya:2001pr}, we consider the scalars $\phi_1$ and $\phi_2$ as homogeneous solutions with classical, local fluctuations. Then we identify the conditions for the exponential growth of these fluctuations, which signals the onset of non-linearities and the formation of Q-balls. The equations of motion for non-homogeneous fields $\Phi_i(x,t)= \phi_i(t)+\delta\phi_i(x,t)$ in an FRW background are given by
\be\begin{split}
&3 H \dot\Phi_1-\Box\Phi_1+f f_1 (\partial\Phi_2)^2 + V_1 = 0 \,,  \\
&3Hf^2\dot\Phi_2 -f^2\Box\Phi_2 - 2 f f_1 \left( \partial\Phi_1 \cdot \partial\Phi_2 \right) = 0 \,,
\end{split}
\ee
where $\Box=-\partial_t^2 + a^{-2} \nabla^2$. For the homogeneous part $\phi_i(t)$ the equations of motions reduce to \eqref{eomphi1phi2}, and for the fluctuations we find
\be\begin{split}
& 3H\delta\dot\phi_1-\Box\delta\phi_1 -\left( f_1^2 + f f_{11}\right) {\dot\phi_2}^2 \delta\phi_1+ V_{11} \delta\phi_1 -2 f f_1 \dot\phi_2\delta\dot\phi_2=0 
 \,,  \\
& 3H f^2\delta\dot\phi_2 - f^2 \Box\delta\phi_2 +2\left(f f_{11}-f_1^2 \right) \dot\phi_1\dot\phi_2\delta\phi_1 
+ 2 f f_1 \left(\dot\phi_2\delta\dot\phi_1  + \dot\phi_1\delta\dot\phi_2 \right) = 0\,.
\end{split}
\ee
The Q-ball forming solutions can be written as\footnote{A more general time and field dependence would not change the main features of the solutions.}
\be
\delta\phi_i=\delta\phi_i^{(0)} e^{\Omega t + i k x}\,,
\ee
with $\Omega \in \mathbb{R}^+$ and $\delta\phi_i^{(0)}\neq0$. The equations of motion for such exponentially growing fluctuations reduce to
\be\begin{split}
& 3H\Omega\delta\phi_1+\left(\Omega^2+\frac{k^2}{a^2}\right)\delta\phi_1 -\left( f_1^2 + f f_{11}\right) {\dot\phi_2}^2 \delta\phi_1+ V_{11} \delta\phi_1 
-2 f f_1 \dot\phi_2\Omega\delta\phi_2 = 0 \,,  \\
& 3H f^2\Omega\delta\phi_2 +  f^2\left(\Omega^2+\frac{k^2}{a^2}\right)\delta\phi_2+2\left(f f_{11}-f_1^2\right) \dot\phi_1\dot\phi_2\delta\phi_1
+ 2 f f_1 \Omega \left(\dot\phi_2 \delta\phi_1  +  \dot\phi_1\delta\phi_2 \right) = 0 \,, 
\end{split}
\ee
which combine to
\be\begin{split}
&\left(3Hf^2\Omega+f^2\left(\Omega^2+\frac{k^2}{a^2}\right) + 2\Omega f f_1\dot\phi_1\right)\left(3H\Omega+\left(\Omega^2+\frac{k^2}{a^2}\right) 
+V_{11} -(f_1^2 + ff_{11})\dot\phi_2^2 \right)  \\
&+ 4\Omega\dot\phi_2^2 f f_1 \left( (f f_{11}-f_1^2)\dot\phi_1 + \Omega f f_1 \right) =0\,.
\label{eqn:qballeom}
\end{split}\ee

\begin{figure}[h!]
	\centering
	\begin{minipage}[b]{0.55\linewidth}
	\centering
	\includegraphics[width=\textwidth]{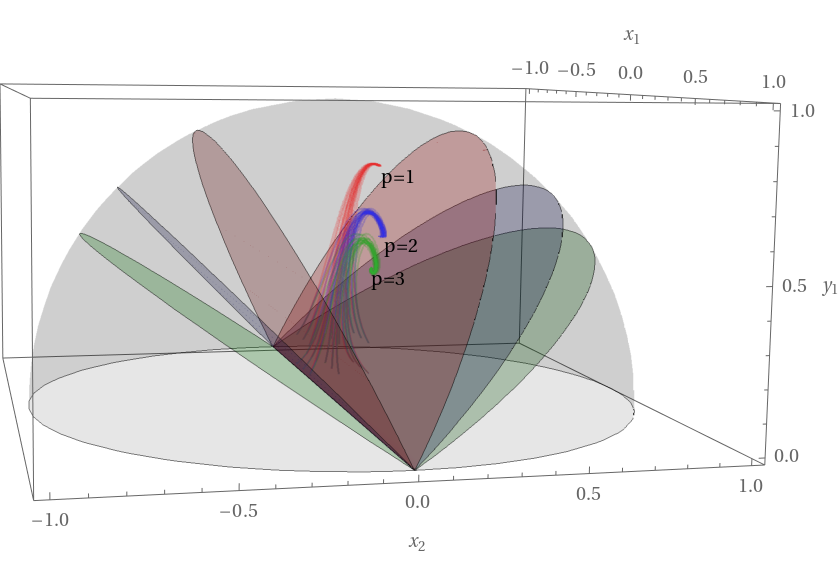}
    \end{minipage}
\caption{Evolution of the two-field quintessence system inside the wedge free of Q-ball formation for universal string moduli and $p=1,2,3$.}
\label{ExamplePlot}
\end{figure}

Assuming $H\simeq 0$ and $\phi_1\simeq\,{\rm const}$, i.e. neglecting the effect of cosmic expansion and assuming a predominantly circular trajectory of the axion, this simplifies to (for $f\neq 0$)
\be 
\left(\Omega^2+\frac{k^2}{a^2}\right)\left(\Omega^2+\frac{k^2}{a^2} +V_{11} -(f_1^2 + f f_{11})\dot\phi_2^2 \right)
+4\Omega^2\dot\phi_2^2 f_1^2  = \Omega^4 + \lambda_2 \Omega^2 + \frac{k^2}{a^2} \lambda_0 =0\,,
\label{p2(O2)}
\ee
where
\be
\lambda_2  \equiv \frac{k^2}{a^2} + 4\dot\phi_2^2 f_1^2 + \lambda_0 \qquad\text{and}\qquad\lambda_0 \equiv \frac{k^2}{a^2} + V_{11} - (f_1^2+ f f_{11})\dot\phi_2^2 \,.
\ee
Clearly, if $\lambda_0\geq 0$, also $\lambda_2 \geq 0$, leading to no solution with $\Omega\in \mathbb{R}^+$. $\lambda_0<0$ is therefore a necessary condition for Q-ball formation which reads
\be
0<\frac{k^2}{a^2} \,<\, (f_1^2 +f f_{11})\dot\phi_2^2 - V_{11} \,,
\label{eqn:qballcondition}
\ee 
where the first inequality just states that the fluctuations are inhomogeneous. Notice that the condition (\ref{eqn:qballcondition}) is also sufficient since the function $\Omega^4 + \lambda_2 \Omega^2 + \frac{k^2}{a^2} \lambda_0$ is positive for $\Omega\to \infty$ while it is negative for $\Omega\to 0$ if $\lambda_0<0$, implying for continuity a value $\Omega\in \mathbb{R}^+$ where it crosses zero. For exponential potentials and kinetic couplings, \eqref{eqn:qballcondition} can be written in terms of $x_2$, $y_1$ and $k_i$ as
\be
0<\frac{k^2}{a^2} \,<\, 3 H^2\,\left( 4 k_1^2x_2^2 - k_2^2 y_1^2 \right),
\label{Wedge}
\ee
leaving a wedge in parameter space which is free of Q-balls. In particular, the dark energy models studied in Sec. \ref{sec:stringy} which involve universal string moduli, seem to have dynamics free of Q-ball formation. Fig. \ref{ExamplePlot} shows that the three trajectories (for $p=1,2,3$) with matter domination initial conditions are predominantly inside the wedge defined by (\ref{Wedge}). Qualitatively similar considerations hold for the three trajectories which pass through today's observations. 

To be somewhat more precise, our matter dominated initial conditions correspond to small perturbations around the unstable $x_1=x_2=y_1=0$ fixed point. In this case the system might initially be in a region in parameter space where Q-ball formation is possible. However we will show now that the Q-ball formation time is much larger than the time for the system to evolve inside the Q-ball free wedge. 

The time scale of Q-ball formation can be estimated by the exponent $\Omega$ of the fluctuations. The roots of the quartic eq. (\ref{p2(O2)}) are given by 
\be
\Omega=\pm \frac{1}{\sqrt{2}}\sqrt{-\lambda_2 \pm\sqrt{\lambda_2^2 + 4\frac{k^2}{a^2} |\lambda_0|}}\,,
\ee
where $\lambda_0=-|\lambda_0|<0$ to allow for Q-ball formation. Since we search for real and positive solutions, we select the `+' signs. As expected from the previous analysis, when $|\lambda_0|\to 0$ the solutions go to zero. On the other hand, for $|\lambda_0|\gg k^2/a^2$, the solutions can be approximated as $\Omega \simeq \sqrt{|\lambda_0|}$. For constant $k_1$ and $k_2$, we can estimate the Q-ball formation time as
\be
\Omega \simeq \sqrt{3}H  \sqrt{4k_1^2x_2^2-k_2^2y_1^2} < 2\sqrt{3}H k_1x_2 \qquad\Rightarrow \qquad
t\sim \frac{1}{\Omega} > \frac{1}{2\sqrt{3} H k_1x_2} \gg \frac{1}{H}\,,
\ee
for small initial values of $x_2$ and $k_1=\mathcal{O}(1)$. Let us stress that these approximations are valid in the physically interesting regime that we choose for initial conditions and parameters.

However the above analysis is not necessarily accurate enough, since we assumed $H\simeq \dot\phi_1 \simeq 0$. Reinstating $H$ and $\dot\phi_1$ using \eqref{eqn:qballeom}, eq. (\ref{p2(O2)}) generalizes to 
\be
\Omega^4 +  \Omega^3 \left(3  H+ \lambda_3 \right)  + \Omega^2 \left(\lambda_2 + 3H \lambda_3 \right) 
+ \Omega \left(3H \frac{k^2}{a^2} +\lambda_3 \lambda_0-\frac{4}{M_p}\, f f_1 \dot\phi_2^2 \dot\phi_1 \partial_{\phi_1} k_1\right) +\frac{k^2}{a^2} \lambda_0=0\,,
\label{Omegaeq}
\ee
where
\be
\lambda_3\equiv 3 H  + 2 \frac{f_1}{f}\dot\phi_1 = H \left(3 - 2\sqrt{6} k_1 x_1\right)\,.
\ee
We have already used $\partial_{\phi_1} k_1=(f_1^2-ff_{11})$ in the linear terms for brevity.
Interestingly, the new terms do not affect the $\Omega\to 0$ and $\Omega\to \infty$ limits, and so $\lambda_0<0$ still remains a sufficient condition. The same condition becomes also necessary if all the coefficients in (\ref{Omegaeq}) are non-negative for $\lambda_0\geq 0$. This is clearly the case if $k_1$ is constant and $\lambda_3\geq 0$, i.e. if $x_1 \leq \sqrt{3/2}/(2 k_1)$. 

Focusing again on the case of two-field quintessence driven by universal string moduli, where $k_1$ is indeed constant and given by (\ref{k1k2fromp}), the condition for Q-ball formation (\ref{eqn:qballcondition}) is definitely necessary for $x_1\leq \sqrt{3p}/4$. This implies that when $x_1>\sqrt{3p}/4$ the conditions for Q-ball formation could be satisfied even inside the wedge defined by (\ref{Wedge}). Let us mention the main features of the trajectories analyzed in Sec. \ref{sec:stringy}
\begin{itemize}
\item For $p=1$, the trajectories with matter dominated initial conditions satisfy the condition $x_1\leq \sqrt{3}/4$ everywhere except for a region close to the fixed point where $x_1 = 1/\sqrt{3} > \sqrt{3}/4$. Hence at the fixed point $\lambda_0>0$ while $\lambda_3<0$, implying that (\ref{Omegaeq}) will definitely have a solution $\Omega \in \mathbb{R}^+$ for $k\to 0$. This signals that the evolution of the system for $p=1$ will end with Q-ball formation in the far future since there is enough time for Q-balls to form close to the fixed point. 

For solutions which pass through today's observations, $x_1\leq \sqrt{3}/4$ also holds everywhere except for two regions: ($i$) close to the fixed point, implying that also these trajectories will end with Q-ball formation; ($ii$) in the vicinity of the putative initial conditions $\omega_{\phi_1}\simeq 0$ and $\Omega_{\phi_1}\simeq 0.55$. However the system will move to $x_1\leq \sqrt{3}/4$ already at $\omega_{\phi_1}\simeq -0.3$ and $\Omega_{\phi_1}\simeq 0.525$, and so before Q-balls have enough time to form. In fact, it can be easily checked that in this region $|\lambda_3|\sim\mathcal{O}(0.1) H$, implying a solution to (\ref{Omegaeq}) which scale as $\Omega \sim \mathcal{O}(0.1)\,H$, and so a very long Q-ball formation time, $t\sim \Omega^{-1}\sim \mathcal{O}(10)/H$. 

\item For $p=2$ and $p=3$, the condition $x_1\leq \sqrt{3p}/4$ is satisfied everywhere along the trajectories, including the fixed points in the far future, except for short transients close to these final fixed points. However, similarly to the $p=1$ case, Q-balls do not have enough time to form since $|\lambda_3|$ remains always small, i.e. $|\lambda_3|\sim\mathcal{O}(0.1) H$. Interestingly, for $p=2$ and $p=3$, $x_1\leq \sqrt{3}/4$ holds around the potential initial conditions, $\omega_{\phi_1}\simeq 0$ and $\Omega_{\phi_1}\simeq 0.6$, for the trajectories which reproduce current data.
\end{itemize}
Summarizing, we have shown that the obstruction of quintessence dynamics by Q-ball formation is never a problem for two-field quintessence models from string theory. Transients where the conditions for exponential growth of non-linearities are satisfied do exist, but they disappear before Q-balls have enough time to form. Interestingly, we found trajectories with $p=1$ which end with Q-ball formation in the far cosmological future.

\section{Conclusions}
\label{Concl}

The mechanism responsible for the present day accelerated expansion of our universe is without any doubt the greatest mystery of fundamental physics. Recent swampland conjectures are expected to shed light on this delicate issue since their goal is to determine the constraints that quantum gravity imposes on any consistent effective field theory. In particular, the dS conjecture \cite{Obied:2018sgi,Garg:2018reu,Ooguri:2018wrx,Andriot:2018mav} would seem to rule out both metastable vacua with a positive cosmological constant and single-field dynamical dark energy models \cite{Cicoli:2021fsd, Cicoli:2021skd} . 

Motivated by these considerations, in this paper we investigated the possibility of realizing multifield quintessence in string theory for two main reasons: ($i$) two-field models with a curved field space have already been shown to be able to lead to late-time accelerating solutions even for steep potentials which are compatible with the dS conjecture \cite{Cicoli:2020cfj,Cicoli:2020noz}; ($ii$) non-linear sigma models arise very naturally in the 4D supergravity effective action of string compactifications. 

We studied two classes of models depending on whether the quintessence complex field is a universal modulus, like the axio-dilaton or the overall K\"ahler modulus, or a non-universal modulus, like a local mode blowing up a Calabi-Yau singularity. While if it is not difficult to find accelerating solutions, we have not been able to obtain multifield trajectories which are compatible with observations both today and in the past. In fact, if the universe matches observations in the past, i.e. the system start from matter domination, there is generically no trajectory which reproduces $\omega_{\phi_1}\simeq -1$ and $\Omega_{\phi_1}\simeq 0.7$ in the future. On the other hand, imposing the system to pass through today's observable universe, the initial epoch turns out to be kinetic domination. 

These somewhat negative results indicate that embedding multifield quintessence in explicit string compactifications seems to be very challenging, especially when compatibility with observations is required. Hence, from this perspective, dynamical dark energy does not seem a preferred alternative to the simplest cosmological constant explanation of the current accelerated expansion of our universe. In this sense, our results might also be seen as raising doubts about the validity of exisiting dS conjectures.

It is important to stress that in this paper we focused just on two-field models, while a generic string compactification might feature many more fields which play an active role during multifield quintessence. While this more involved scenario would definitely need a more accurate analysis, we expect our results to still hold qualitatively since the two-field dynamics with kinetic couplings should already catch the main features of the departure from a simple single-field geodesic dynamics.

Let us finally point out that the models we have considered feature a conserved charge due to the underlying axionic shift symmetry. This can lead to an exponential growth of non-linearities that can end up in Q-ball formation and a consequent sudden obstraction of the dark energy dynamics. However we have shown that our models are free from Q-ball formation during their entire cosmological evolution except for some cases where Q-balls might form only in the far future when the system approaches a fixed point.

\clearpage
\bibliography{references}  
\bibliographystyle{utphys}

\end{document}